\documentclass[preprint,3p,times,twocolumn]{elsarticle}
\usepackage{ecrc}
\usepackage{hyperref}
\usepackage{amssymb}
\usepackage[figuresright]{rotating}
\usepackage{latexsym}
\usepackage{amsmath,amssymb,amsfonts,wasysym}
\usepackage{graphics}
\usepackage{xcolor}
\usepackage{float}
\usepackage{hyperref}
\usepackage{longtable}
\usepackage{verbatim}
\volume{00}
\firstpage{1}
\journalname{Nuclear Physics B Proceedings Supplement}
\runauth{}
\jid{nuphbp}
\jnltitlelogo{Nuclear Physics B Proceedings Supplement}
\input{tcilatex}
\begin{document}

\begin{frontmatter}



\dochead{}

\title{Composite Resonances effects on EWPT and Higgs diphoton decay rate}


\author{A. E. C\'arcamo Hern\'andez, Claudio O. Dib and Alfonso R. Zerwekh}
\runauth{A. E. C\'arcamo Hern\'andez, Claudio O. Dib and Alfonso R. Zerwekh}

\address{Universidad T\'ecnica Federico Santa Mar\'{\i}a and Centro Cient%
\'{\i}fico-Tecnol\'ogico de Valpara\'{\i}so, Casilla 110-V, Valpara\'{\i}so,
Chile}

\begin{abstract}
In scenarios of strongly coupled electroweak symmetry breaking,
heavy composite particles of different spin and parity may arise and cause
observable effects on signals that appear at loop levels. The recently observed process of Higgs to $\gamma \gamma$ at the LHC is one of such signals. We study the new constraints that 
are imposed on composite models from  $H\to \gamma\gamma$, together with the existing constraints from the high precision electroweak tests.  We use an effective chiral Lagrangian to
describe the effective theory that contains the Standard Model spectrum and the extra composites
below the electroweak scale. 
Considering the effective theory  cutoff at $\Lambda = 4\pi v \sim 3 $ TeV,  consistency with
the $T$ and $S$ parameters and the newly observed $H\to \gamma
\gamma$  can be found for a rather restricted range of masses of vector and axial-vector composites from $1.5$ TeV to $1.7$ TeV and $1.8$ TeV to $1.9$ TeV, respectively, and only 
provided a non-standard kinetic mixing between the $W^{3}$ and $B^{0}$ fields is
included.
\end{abstract}

\begin{keyword}
Composite Higgs Models, Composite spin-1 and spin-0 resonances, EWT, Diphoton decay rate.


\end{keyword}

\end{frontmatter}

% \linenumbers


\section{Introduction}

\label{intro}

One of the possible signals of composite Higgs boson models is the deviation
of the $h\rightarrow \gamma \gamma$ channel from the Standard Model (SM)
prediction, as it is a loop process sensitive to heavier virtual states. For
instance this signal was predicted in the context of Minimal Walking
Technicolor \cite{Hapola:2011sd}. Consequently the recent $h\rightarrow
\gamma \gamma$ signal reported by ATLAS and CMS collaborations \cite{atlashiggs,cmshiggs,newtevatron,CMS-PAS-HIG-12-020}, which is very close to
the SM prediction, implies an additional constraint on composite models. In
this regard, it is important to explore the consequences of this new
constraint on composite models, in conjunction with those previously known
from electroweak precision measurements.

Given the recent evidence of the Higgs boson, a strongly interacting sector
that is phenomenologically viable nowadays should include this scalar boson
in its low energy spectrum, but it is also assumed that vector and
axial-vector resonances should appear as well, in a way that the so called
Weinberg sum rules \cite{WeinbergSumRules} are satisfied \cite{Appelquist:1998xf,Foadi:2007ue,Foadi:2007se}.

Here we formulate this kind of scenario in a general way, without referring
to the details of the underlying strong dynamics, by using a low energy
effective Lagrangian which incorporates vector and axial-vector resonances,
as well as composite scalars. One of these scalars should be the observed
Higgs and the others should be heavier as to avoid detection at the LHC. Our
inclusion of the vector and axial resonances is based on a 4-site Hidden
Local Symmetry, which requires three scalar sectors (link fields)
responsible for the breaking of the hidden local symmetries. This setup
naturally leads to a spectrum that contains three physical scalars.

The main reason to still consider strongly interacting mechanisms of
electroweak symmetry breaking (EWSB) as alternatives to the Standard Model
mechanism is the so called hierarchy problem that arises from the Higgs
sector of the SM. This problem is indicative that, in a natural scenario,
new physics should appear at scales not much higher than the EWSB scale
(say, around a few TeV) in order to stabilize the Higgs mass at scales much
lower than the Planck scale ($\sim 10^{19}$ GeV). An underlying strongly
interacting dynamics without fundamental scalars, which becomes
non-perturbative somewhere above the EW scale, is a possible scenario that
gives an answer to this problem. The strong dynamics causes the breakdown of
the electroweak symmetry through the formation of condensates in the vacuum 
\cite{Appelquist:2003,Hirn:2007,Hirn:2008tc,Belyaev:2008yj,Hill:2003}.

%A 125 GeV composite Higgs boson versus flavour and electroweak precision tests
Many models of strong EWSB have been proposed which predict the existence of
composite particles such as scalars \cite{ggpr,Barbieri:2007,Burdman:2007,Lodone:2008,Zerwekh:2010,Burdman:2011,Carcamo:2012,Pomarol:2012,Barbieri:2012,Foadi:2012bb,Castillo-Felisola:2013jva,Moretti:2013,Pappadopulo:2013vca,Contino:2013kra}%
, vectors \cite{Bagger,Casalbuoni:1995,SekharChivukula:2001,Csaki:2003,Barbieri:2008,Barbieri:2010,Carcamo:2010a}%
, both scalars and vectors \cite{Casalbuoni:1985,Casalbuoni:1986vq,Casalbuoni:1988xm,Dominici:1997zh,Zerwekh:2006,Carcamo:2010,Carcamo:ggtoVV,Carcamo:2011,Bellazini:2012,Contino:2012,Carcamo-Hernandez:2013ypa,Castillo-Felisola:2013jua,Pappadopulo:2014qza,Barbieri:2015lqa,Low:2015nqa,Hernandez:2015nga}
and composite fermions \cite{Kaplan:1991dc,Barbieri:2008b}. These predicted
scalar and vector resonances play a very important role in preserving the
unitarity of longitudinal gauge boson scattering up to the cutoff $\Lambda
\simeq 4\pi v$ \cite{Barbieri:2003pr,Nomura2003,Foadi:2003xa,Georgi:2004iy,SekharChivukula:2008mj,Sanino:2008}%
. One should add that a composite scalar does not have the hierarchy problem
since quantum corrections to its mass are cut off at the compositeness
scale, which is assumed to be much lower than the Planck scale.

In this work we assume that Electroweak Symmetry Breaking is due to an
underlying strongly interacting sector that possesses a global $%
SU(2)_{L}\times SU(2)_{R}$ symmetry, which breaks down to the subgroup $%
SU(2)_{L+R}$. The SM electroweak symmetry $SU(2)_L\times U(1)_Y$ is assumed
to be embedded as a local part of the $SU(2)_{L}\times SU(2)_{R}$ symmetry,
so the spontaneous breaking of the latter leads to EWSB. The strong dynamics
responsible for EWSB in general gives rise to massive composite fields. We
will assume that only spin-zero and spin-one composites are lighter than the cutoff $\Lambda \simeq 4\pi v$ so that they explicitly appear as fields in
the effective chiral Lagrangian. Composite states of spin 2 and higher are assumed to be heavier than the cutoff, and so are disregarded in this work. 
Consequently, the spectrum below the cutoff will have vector and axial vector fields ($V_\mu^a$ and $A_\mu^a$, respectively) belonging to the triplet representation of the $%
SU(2)_{L+R}$ custodial group, as well as two massive composite scalars ($h$
and $H$) and one pseudoscalar ($\eta$), all singlets under that group. We
will identify the lightest scalar, $h$, with the state of mass $m_{h}=126$
GeV discovered at the LHC. Concerning the coupling to fermions, the spin-one
fields $V_{\mu }^{a}$ and $A_{\mu }^{a}$ will couple to SM fermions only
through their kinetic mixings with the SM Gauge bosons, and the spin zero
fields $h$, $H$ and $\eta$ interact with the fermions only via
(proto)-Yukawa couplings.

%In this work we assume a scenario where there is a strongly interacting sector which possesses a global $SU(2)_{L}\times SU(2)_{R}$ symmetry. The strong dynamics responsible for EWSB, in general gives rise to massive composite vector and axial vector fields ($V_\mu^a$ and $A_\mu^a$% , respectively) belonging to the triplet representation of the $SU(2)_{L+R}$ custodial group, as well as two massive composite scalars ($h$ and $H$) and one pseudoscalar ($\eta$) all singlets under that group. We will identify the lightest scalar, $h$, with the state of mass $m_{h}=126$ GeV discovered at the LHC. All of these composite resonances are assumed to be lighter than the cutoff $\Lambda \simeq 4\pi v$, so that they explicitly appear as fields in the effective chiral Lagrangian. In this work we use the high precision results on $S$ and $T$ and the recent ATLAS and CMS results at the LHC on $h\to\gamma\gamma$ to constrain the mass and coupling parameters of the model.\newline

In this work, we build an effective chiral lagrangian to represent this
generic scenario below the symmetry breaking cutoff and study its
consistency with the current phenomenology. In particular we study the
effects on the high precision results on $S$ and $T$ and the recent ATLAS
and CMS results at the LHC on $h\to\gamma\gamma$, all of which are loop
processes that are sensitive to heavy virtual particles.

\section{The Model}

%The Lagrangian of our effective theory is given by \cite% {Carcamo-Hernandez:2013ypa}: 
We formulate our strongly coupled sector by means of an
effective chiral Lagrangian that incorporates the heavy composite states by
means of local hidden symmetries \cite{Bando:1985rf}. 
As shown in detail Ref. \cite{Carcamo-Hernandez:2013ypa}, this Lagrangian is based on the symmetry $G=SU\left( 2\right) _{L}\times
SU\left( 2\right)_{C}\times SU\left( 2\right) _{D}\times SU\left( 2\right)_{R}$. The $SU\left( 2\right)_{C}\times SU\left( 2\right) _{D}$ part is a
hidden local symmetry whose gauge bosons are linear combinations of the vector and axial-vector composites, and the SM gauge fields (see Ref. \cite{Carcamo-Hernandez:2013ypa} for details). The SM gauge group, on the other hand, is contained as a local form of the $SU\left( 2\right)_{L}\times SU\left( 2\right) _{R}$ global symmetry of the underlying dynamics.

As the symmetry $G$ is spontaneously broken down to the diagonal subgroup $%
SU(2)_{L+C+D+R}$, it is realized in a non-linear way with the inclusion of
three link fields (spin-0 multiplets). These link fields contain two
physical scalars $h$ and $H$, one physical pseudoscalar $\eta$, the three
would-be Goldstone bosons absorbed as longitudinal modes of the SM gauge
fields and the six would-be Goldstone bosons absorbed by the composite
triplets $V_\mu$ and $A_\mu$. In the framework of strongly interacting
dynamics for EWSB, the interactions below the EWSB
scale among the SM particles and the extra composites can be described by the effective Lagrangian \cite{Carcamo-Hernandez:2013ypa}:
\begin{eqnarray}
\mathcal{L} &=&\frac{v^{2}}{4}\left\langle D_{\mu }UD^{\mu }U^{\dagger
}\right\rangle -\frac{1}{2g^{2}}\left\langle W_{\mu \nu }W^{\mu \nu
}\right\rangle   \notag \\
&&-\frac{1}{2g^{\prime 2}}\left\langle B_{\mu \nu }B^{\mu \nu }\right\rangle
+\frac{c_{WB}}{4}\left\langle U^{\dagger }W_{\mu \nu }UB^{\mu \nu
}\right\rangle   \notag \\
&&+\sum_{R=V,A}\left[ -\frac{1}{4}\left\langle R_{\mu \nu }R^{\mu \nu
}\right\rangle +\frac{1}{2}M_{R}^{2}R_{\mu }R^{\mu }\right]   \notag \\
&&-\frac{f_{V}}{2\sqrt{2}}\left\langle V^{\mu \nu }\left( uW_{\mu \nu
}u^{\dagger }+u^{\dagger }B_{\mu \nu }u\right) \right\rangle   \notag \\
&&-\frac{ig_{V}}{2\sqrt{2}}\left\langle V^{\mu \nu }\left[ u_{\mu },u_{\nu }%
\right] \right\rangle +\frac{f_{A}}{2\sqrt{2}}\left\langle u_{\mu \nu
}A^{\mu \nu }\right\rangle   \notag \\
&&-\frac{if_{A}}{2\sqrt{2}}\left\langle \left( uW_{\mu \nu }u^{\dagger
}+u^{\dagger }B_{\mu \nu }u\right) \left[ A^{\mu },u^{\nu }\right]
\right\rangle   \notag \\
&&-\frac{i\kappa f_{A}}{2\sqrt{2}}\left\langle u_{\mu \nu }\left[ V^{\mu
},u^{\nu }\right] \right\rangle +\alpha _{\eta }\left\langle V_{\mu }A^{\mu
}\right\rangle \eta   \notag \\
&&+\beta _{\eta }\left\langle V_{\mu }u^{\mu }\right\rangle \eta +\mathcal{L^{\prime }}\notag\\
&&+\sum_{S=h,H,\eta }\left[ \frac{1}{2}\partial _{\mu }S\partial ^{\mu }S+%
\frac{m_{S}^{2}}{2}S^{2}\right]   \notag \\
&&+\sum_{S=h,H}\left[ \alpha _{S}S\left\langle u_{\mu }u^{\mu }\right\rangle
+\beta _{S}S\left\langle V_{\mu }V^{\mu }\right\rangle \right]  \notag \\
&&+\sum_{S=h,H}\left[ \gamma _{S}S\left\langle A_{\mu }A^{\mu }\right\rangle
+\delta _{S}S\left\langle A_{\mu }u^{\mu }\right\rangle \right] 
\label{Lagrangian}
\end{eqnarray}%
where $\mathcal{L^{\prime }}$ corresponds to the part of the Lagrangian
which includes: the interactions of \emph{two} of the heavy spin-one fields
with the SM Goldstone bosons and gauge fields, the interactions involving 
\emph{three} heavy spin-one fields, the quartic self-interactions of $V_{\mu
}$ and of $A_{\mu }$, the contact interactions involving the SM gauge fields
and Goldstone bosons, the interaction terms that include two of the
spin-zero fields coupled to the SM Goldstone bosons or gauge fields, or to
the composite $V_{\mu }$ and $A_{\mu }$ fields, the mass terms for the SM
quarks as well as interactions between the spin-0 fields $h$, $H$ and $\eta $
and the SM fermions. Besides that, the dimensionless couplings in Eq. (\ref%
{Lagrangian}) are given in Ref.\cite{Carcamo-Hernandez:2013ypa}, and the
following definitions are fulfilled: 
\begin{equation}
\begin{array}{l}
U\left( x\right) =e^{i\hat{\pi}\left( x\right) /v}\,,\qquad \hat{\pi}\left(
x\right) =\tau ^{a}\pi ^{a}\,,\qquad u\equiv \sqrt{U},\qquad  \\ 
{B}_{\mu }=\frac{g^{\prime }}{{2}}\tau ^{3}B_{\mu }^{0},\ ,\qquad {W}_{\mu }=%
\frac{g}{{2}}\tau ^{a}W_{\mu }^{a}\,,\qquad R_{\mu }=\frac{1}{\sqrt{2}}\tau
^{a}R_{\mu }^{a},\\ 
R=V,A\,,\qquad D_{\mu }U=\partial _{\mu }U-i{B}_{\mu }U+iU{W}_{\mu }\,, \\ 
\hat{X}_{\mu \nu }=\nabla _{\mu }X_{\nu }-\nabla _{\nu }X_{\mu },\qquad
X=R,u\qquad u_{\mu }=iu^{\dag }D_{\mu }Uu^{\dag }, \\ 
\nabla _{\mu }{R}=\partial _{\mu }{R}+[\Gamma _{\mu },{R}]\,,\qquad  \\ 
\Gamma _{\mu }=\frac{1}{2}\Big[u^{\dag }\left( \partial _{\mu }-i{B}_{\mu
}\right) u+u\left( \partial _{\mu }-i{W}_{\mu }\right) u^{\dag }\Big].%
\end{array}%
\end{equation}%

Our effective theory is based on the following assumptions \cite{Carcamo-Hernandez:2013ypa}:

\begin{enumerate}
\item The Lagrangian responsible for EWSB has an underlying strong dynamics
with a global $SU(2)_{L}\times SU(2) _{R}$ symmetry which is spontaneously
broken by the strong dynamics down to the $SU(2)_{L+R}$ custodial group. The
SM electroweak gauge symmetry $SU(2)_L\times U(1)_Y$ is assumed to be
embedded as a local part of the $SU(2)_{L}\times SU(2) _{R}$ symmetry. Thus
the spontaneous breaking of $SU(2)_{L}\times SU(2) _{R}$ also leads to the
breaking of the electroweak gauge symmetry down to $U(1)_{em}$.

\item The strong dynamics produces composite heavy vector fields $V_{\mu
}^{a}$ and axial vector fields $A_{\mu }^{a}$, triplets under the custodial $%
SU(2)_{L+R} $, as well as a composite scalar singlet $h$ with mass $%
m_{h}=126 $ GeV, a heavier scalar singlet $H$, and a heavier pseudoscalar
singlet $\eta$. These fields are assumed to be the only composites lighter
than the symmetry breaking cutoff $\Lambda \simeq 4\pi v$.

\item The heavy fields $V_{\mu }^{a}$ and $A_{\mu }^{a}$ couple to SM
fermions only through their kinetic mixings with the SM Gauge bosons.

\item The spin zero fields $h$, $H$ and $\eta$ interact with the fermions
only via (proto)-Yukawa couplings.
\end{enumerate}

\section{Study of effects on $T$, $S$ and $h\rightarrow \protect%
\gamma \protect\gamma $.}

In the Standard Model, the $h\rightarrow \gamma \gamma $ decay is dominated
by $W$ loop diagrams which can interfere destructively with the subdominant
top quark loop. In our strongly coupled model, the $h\rightarrow \gamma
\gamma $ decay receives extra contributions from loops with charged $V_{\mu }
$ and $A_{\mu }$, as shown in Figure 1  \cite{Carcamo-Hernandez:2013ypa}. 
\begin{figure}[tbh]
\includegraphics[width=9cm,height=12cm,angle=0]{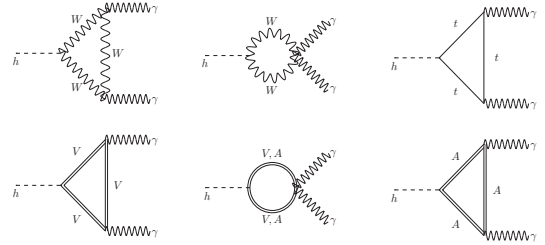}%
\vspace{-7.7cm}
\caption{One loop Feynman diagrams in the Unitary Gauge contributing to the $%
h\rightarrow \protect\gamma \protect\gamma $ decay.}
\label{hto2photons}
\end{figure}%\newline
\begin{figure}[tbh]
\includegraphics[width=9cm,height=12cm,angle=0]{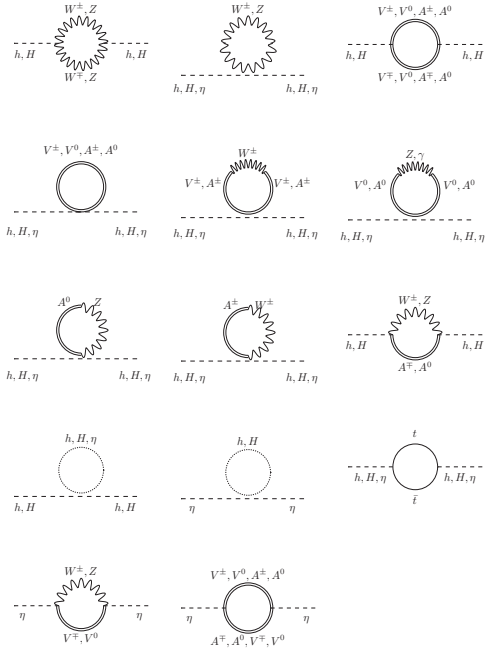}\vspace{%
-2.5cm}
\caption{One loop Feynman diagrams in the Unitary Gauge contributing to the
masses of the parity even $h$ and $H$ and parity odd $\protect\eta $ scalars \cite{Carcamo-Hernandez:2013ypa}.
}
\label{spin0masses}
\end{figure}
Notice that we have not considered the contribution from contact interactions of gluons, such as
\begin{equation}
\tciLaplace _{ggVV}%\subset
=\frac{a_{ggVV}}{\Lambda ^{2}}G_{\mu \nu }G^{\mu
\nu }V_{\alpha }V^{\alpha }.
\end{equation}
to the Higgs production mechanism at the LHC, $gg\to h$, which could have a sizable effect that might contradict the current experiments. Nevertheless, we have checked that this contribution is negligible provided the effective coupling $a_{ggVV}<0.5$. We recall that the heavy vector and heavy axial-vector resonances are colorless, and therefore they do not have renormalizable interactions with gluons.%
In this work we want to determine the range of the heavy vector masses which
is consistent with the events in the $h\rightarrow \gamma \gamma $ decay
recently observed at the LHC. To this end, we will introduce the ratio $%
R_{\gamma \gamma }$, which measures the $\gamma \gamma $ signal produced in
our model relative to the signal within the SM: 
\begin{eqnarray}
R_{\gamma \gamma } &=&\frac{\sigma \left( pp\rightarrow h\right) \Gamma
\left( h\rightarrow \gamma \gamma \right) }{\sigma \left( pp\rightarrow
h\right) _{SM}\Gamma \left( h\rightarrow \gamma \gamma \right) _{SM}}
\label{R_gamma} \\
&\simeq &a_{htt}^{2}\frac{\Gamma \left( h\rightarrow \gamma \gamma \right) }{%
\Gamma \left( h\rightarrow \gamma \gamma \right) _{SM}}.  \notag
\end{eqnarray}%
where $a_{htt}$ is the deviation of the Higgs-top quark coupling with
respect to the SM.\newline
Let us first study the masses of $h$, $H$ and $\eta $ up to one loop. The
one-loop diagrams are shown in Fig.~\ref{spin0masses}. Now, we want $h$ to
be the recently discovered Higgs boson of mass $\sim 126$ GeV, while $H$ and 
$\eta $ should be heavier, their masses satisfying the experimental bound $%
600$ GeV $\lesssim $ $m_{H},m_{\eta }$ $\lesssim $ $1$~TeV. These masses have tree-level contributions directly from the scalar potential, but also important one-loop contributions from the
Feynman diagrams shown in Fig.~\ref{spin0masses}. All these one-loop diagrams have quadratic and
some have also quartic sensitivity to the ultraviolet cutoff $\Lambda$ of the effective theory. 
The calculation details are included in Ref. \cite{Carcamo-Hernandez:2013ypa}. 
As shown there, the contact interaction diagrams
involving $V_\mu$ and $A_\mu$ in the internal lines interfere destructively with
those involving trilinear couplings between the heavy spin-0 and spin-1
bosons. As shown in Ref. \cite{Carcamo-Hernandez:2013ypa}, the quartic couplings of a pair of spin-1 fields with two $h$'s are equal to those with two $H$'s. This implies that contact interactions contribute at one-loop level equally to the $h$ and $H$ masses. On the other hand, since the couplings of two spin-1 fields with one $h$ or one $H$ are different, i.e., $a_{hWW}\neq a_{HWW}$, $a_{hAA}\neq a_{HAA}$, $a_{hWA}\neq a_{HWA}$, $a_{hZA}\neq a_{HZA}$, these loop contributions cause the masses $m_h$ and $m_H$ to be significantly different,  the former being much smaller than the latter 
 (notice that in the Standard Model, $a_{hWW}=b_{hhWW}=1$, implying an exact cancelation of the quartic divergences in the one-loop contributions to the Higgs mass). 
As it turns out, one can easily find conditions where the terms that are quartic in the cutoff cause partial cancelations in $m_h$, but not so in $m_H$ and $m_\eta$, making $m_h$ much lighter that the cutoff $\Lambda$ (e.g. $m_h\sim$ 126 GeV) while $m_H$ and $m_\eta$ remain heavy. 
In Figs.\ \ref{fig0}.a and \ref{fig0}.b we show the sensitivity of the light scalar mass $m_h$ to variations of $M_V$ and $a_{htt}$, respectively. These Figures show that the values of $M_V$ and $a_{htt}$ have an important effect on $m_h$. We can see that these models with composite vectors and axial vectors have the potential to generate scalar masses well below the supposed value around the cutoff, but only in a rather restricted range of parameters. The high sensitivity to the parameters, however, does not exhibit a fine tuning in the usual sense: that deviations from the adjusted point would always bring the mass back to a ``naturally high'' value near the cutoff. Here, the adjustment of parameters could bring the light scalar mass either back up or further below the actual value of 126 GeV  \cite{Carcamo-Hernandez:2013ypa}.%
Let us now analyze the constraints imposed on the parameters by the values
of $T$ and $S$ given by the experimental high precision tests of electroweak interactions. The Feynmann diagrams contributing to the $T$ and $S$
parameters are shown in Figures 3 and 4, respectively. As shown in Ref. \cite{Carcamo-Hernandez:2013ypa}, in general the expressions for $T$ and $S$ exhibit quartic, quadratic and logarithmic dependence on the cutoff $\Lambda \sim 3$ TeV. However, the contributions coming from loops containing the $h$, $H$ and $\eta $ scalars are not very sensitive with the cutoff, as they do not contain quartic terms in $\Lambda$. As a consequence, $T$ and $S$ happen to have a rather mild sensitivity to the masses of $%
H$ and $\eta $, and so we will restrict our study to a scenario where $H$
and $\eta $ are degenerate in mass at a value of $1$ TeV. In contrast, most
of the other diagrams, i.e. those containing SM bosons and/or the composite
spin-1 fields $V_{\mu }$ or $A_{\mu }$, have quartic and quadratic
dependence on the cutoff, and as a consequence they are very sensitive to
the masses $M_{V}$ and $M_{A}$ \cite{Carcamo-Hernandez:2013ypa}.

%The definition of the $T$ and $S$ parameters and their calculation within our model are given in Appendix \ref{A2} [see Eqs.\ (\ref{T}) and (\ref%{Sparameter})]. 
We can separate the contributions to $T$ and $S$ as $T=T_{SM}+\Delta T$ and $%
S=S_{SM}+\Delta S$, where 
\begin{equation}
T_{SM}=-\frac{3}{16\pi \cos ^{2}\theta _{W}}\ln \left( \frac{m_{h}^{2}}{%
m_{W}^{2}}\right) ,\ S_{SM}=\frac{1}{12\pi }\ln \left( \frac{m_{h}^{2}}{%
m_{W}^{2}}\right) 
\end{equation}%
are the contributions within the SM, while $\Delta T$ and $\Delta S$ contain
all the contributions involving the extra particles.
\begin{figure*}[tbh]
\resizebox{0.9\textwidth}{!}{
\includegraphics{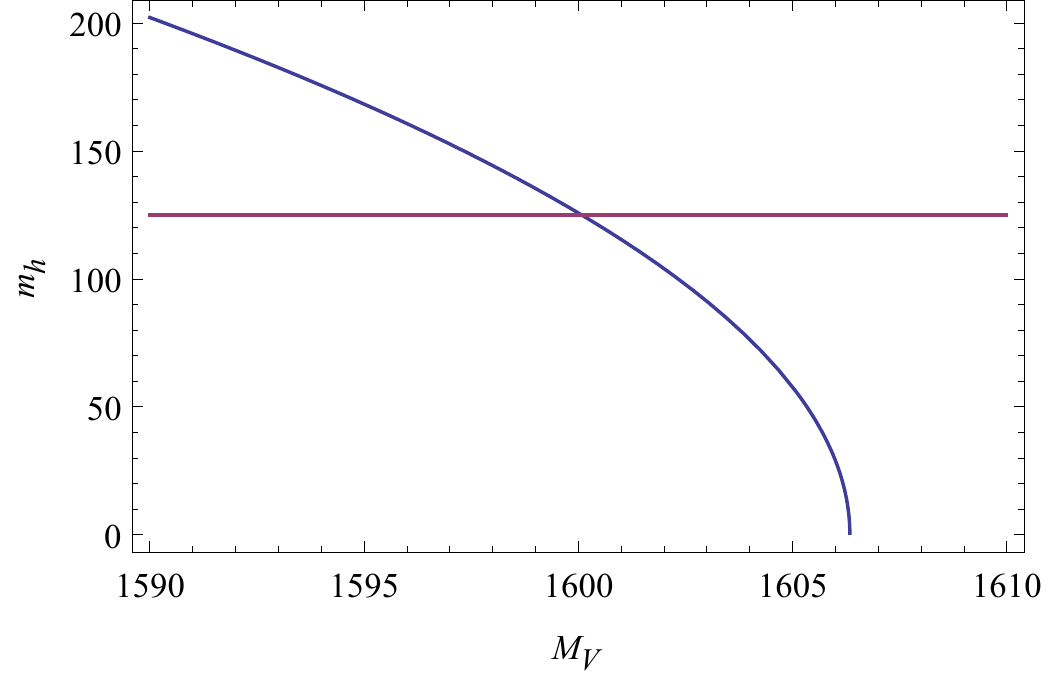}\hspace{2cm}
\includegraphics{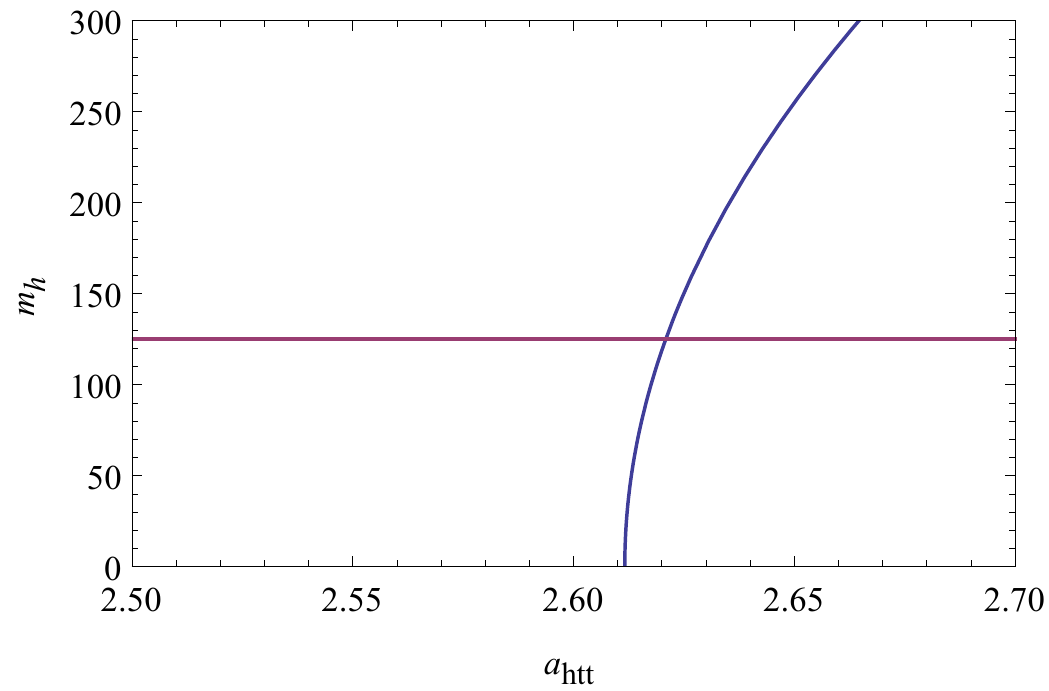}
}
\begin{tabular}{cccc}
\hspace{2cm} &
$\kappa=0.76$, $a_{htt}=2.62$ &
\hspace{3.2cm} &
$\kappa=0.76$, $M_{V}=1.6$ TeV\\
\hspace{2cm} &
(\ref{fig0}.a) &  
\hspace{2cm} &
(\ref{fig0}.b)
\end{tabular}
\caption{Light scalar mass $m_h$ as function of $M_V$ for $\kappa=0.76$, $a_{htt}=2.62$ TeV  (Fig.\ \protect\ref{fig0}.a), $a_{htt}$ for $\kappa=0.76$, $M_{V}=1.6$ TeV (Fig.\ \protect\ref{fig0}.b)  \cite{Carcamo-Hernandez:2013ypa}. The horizontal line corresponds to the value $126$ GeV for the light Higgs boson mass.}
\label{fig0}
\end{figure*} 
\begin{figure}[tbh]
\hspace{-1cm}\includegraphics[width=10cm,height=12cm,angle=0]{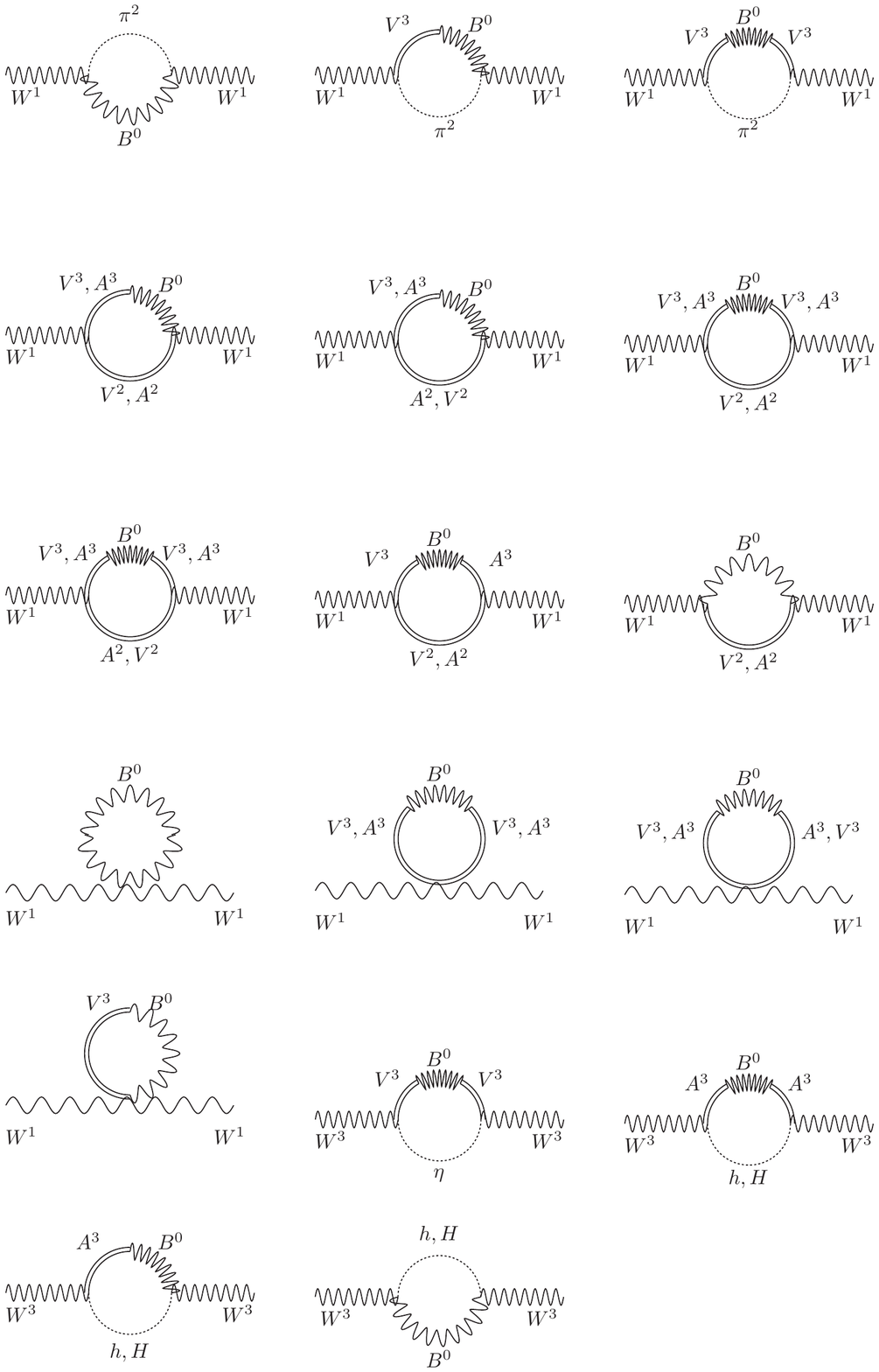}\vspace{-1.3cm}
\caption{One loop Feynman diagrams contributing to the $T$ parameter \cite{Carcamo-Hernandez:2013ypa}.}
\label{figT}
\end{figure}\newline
\begin{figure}[tbh]
\includegraphics[width=8cm,height=12cm,angle=0]{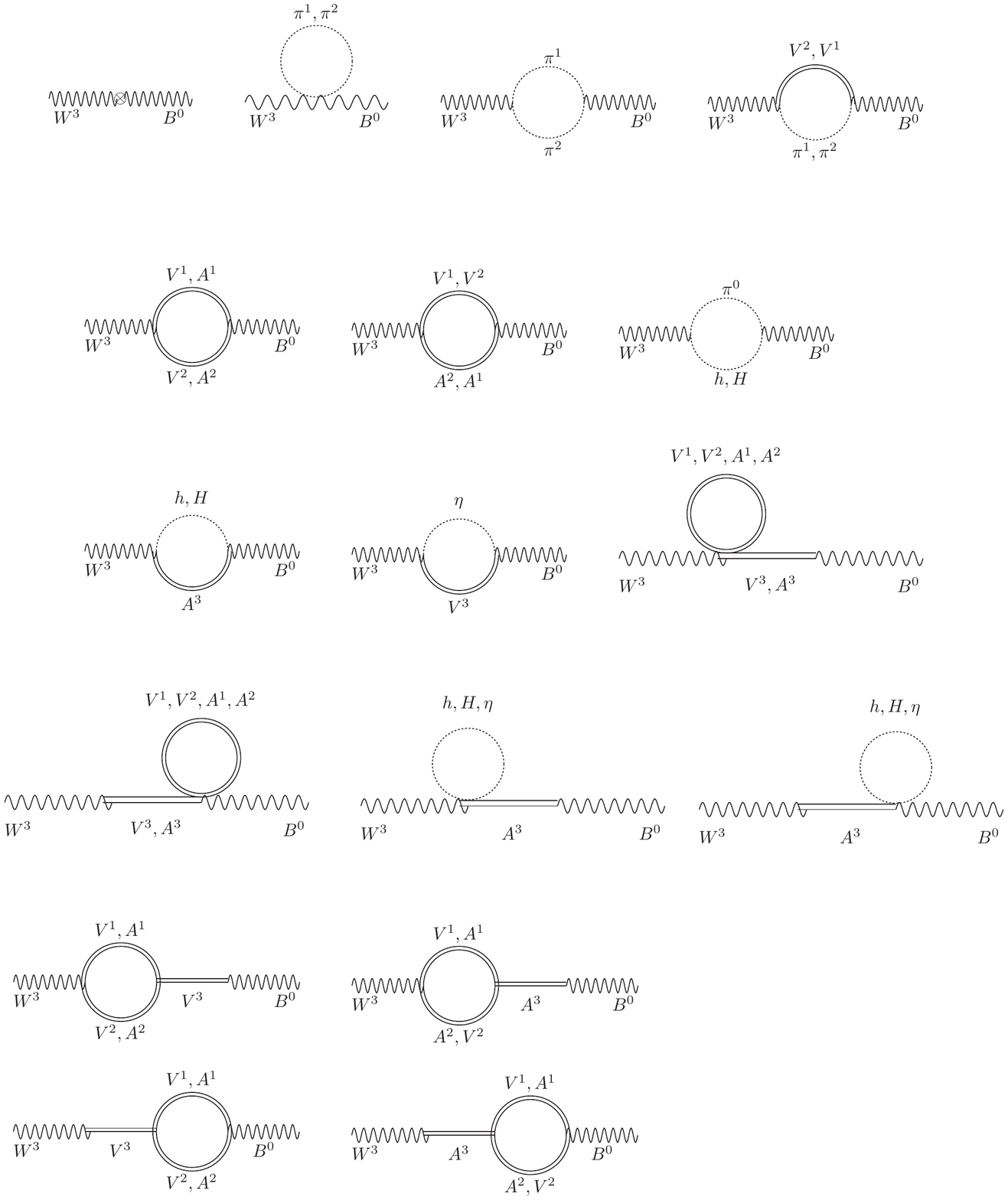}\vspace{-1.4cm}
\caption{One loop Feynman diagrams contributing to the $S$ parameter \cite{Carcamo-Hernandez:2013ypa}.}
\label{figS}
\end{figure}
%
%
%Then, the experimental results on $T$ and $S$ impose the restriction that  $\Delta T$ and $\Delta S$  must lie inside a region in the $\Delta S-\Delta T$ plane. Explicitly, the experimentally allowed region at the $95\%$CL in the $\Delta S-\Delta T$ plane is the ellipse shown in Figs. 5. 
\begin{figure}[tbh]
\centering\includegraphics[width=6.5cm,height=4.5cm,angle=0]{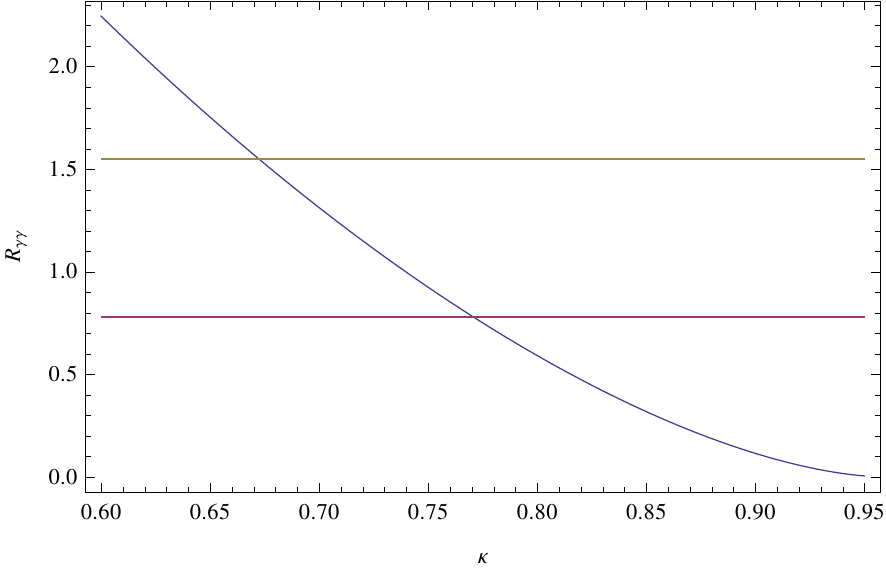} 
%\newline
\caption{The ratio $R_{\protect\gamma \protect\gamma }$ as a function of $%
\protect\kappa $ for $g_{C}v=0.8$ TeV. The horizontal lines are the $R_{%
\protect\gamma \protect\gamma }$ experimental values given by CMS and ATLAS,
which are equal to $1.6\pm 0.4$ and $1.8\pm _{0.419}^{0.460}$, respectively 
\protect\cite{ATLAS-2013,CMS-2013}.}
\label{figR}
\end{figure}
The experimental results on $T$ and $S$ restrict $\Delta T$ and $\Delta S$
to lie inside a region in the $\Delta S-\Delta T$ plane. At the $95\%$
confident level (CL), these regions are the ellipes shown in Figs. 5. We can
now study the restrictions on $a_{htt}$, $M_{V}$ and $\kappa $ imposed by a
mass $m_{h}=125.5$ GeV\ for the light Higgs boson and the two-photon signal $%
0.78\lesssim R_{\gamma \gamma }\lesssim 1.55$, which at the same time
respect the previously described bounds imposed by the $T$ and $S$
parameters at $95\%$ CL. After scanning the parameter space we find that the
heavy vector mass has to be in the range $1.51$ TeV$\lesssim M_{V}\lesssim
1.75$ TeV in order for the $T$ parameter to be consistent with the
experimental data at $95\%$ CL. Regarding the mass ratio $\kappa
=M_{V}^{2}/M_{A}^{2}$ and the Higgs-top coupling $a_{htt}$, we find that
they have to be in the ranges $0.75\lesssim \kappa \lesssim 0.78$ and $%
2.53\lesssim a_{htt}\lesssim 2.72$, respectively. Therefore, the light $126$
GeV Higgs boson in this model couples strongly with the top quark, yet
without spoiling the perturbative regime in the sense that the condition $%
\frac{a_{htt}^{2}}{4\pi }\lesssim 1$ is still fulfilled. Concerning the top
coupling to the heavy pseudoscalar $\eta $, by imposing the experimental
bound $600$ GeV $\lesssim m_{\eta }\lesssim 1$ TeV for heavy spin-0
particles, we find that the coupling has the bound $a_{\eta tt}\lesssim 1.39$
for $M_{V}\simeq 1.51$ TeV, $\kappa \simeq 0.75$ (lower bounds), and $%
a_{\eta tt}\lesssim 1.46$ for $M_{V}\simeq 1.75$ TeV, $\kappa \simeq 0.78$
(upper bounds). Regarding the top coupling to the heavy scalar $H$, we find
that it grows with $m_{H}$, and at the lower bound $m_{H}\sim 600$ GeV the
coupling is restricted to be $a_{Htt}\simeq 3.53$, which implies that $H$
also couples strongly to the top quark. %
Let us now study the restrictions imposed by the two-photon signal, given in
terms of the ratio $R_{\gamma \gamma }$ of Eq.~(\ref{R_gamma}). We explored
the parameter space of $M_{V}$ and $\kappa $ ($\kappa =M_{V}^{2}/M_{A}^{2}$)
trying to find values for $R_{\gamma \gamma }$ within a range more or less
consistent with the ATLAS and CMS results. In Fig.~\ref{figR} we show $%
R_{\gamma \gamma }$ as a function of $\kappa $, for the fixed values $%
g_{C}v=0.8$ TeV and $a_{htt}=2.6$. We recall that $M_V=g_Cv/\sqrt{1-\kappa}$, being $g_C$ the coupling constant of the strong sector. We chose $a_{htt}=2.6$, which is near the
center of the range $2.53\lesssim a_{htt}\lesssim 2.72$ imposed by a light
Higgs boson mass of $m_{h}=125.5$ GeV, as previously described. In turn, 
%Recall that light $125.5$ GeV light Higgs boson strongly couples with the top quark with the corresponding coupling satisfying the bound $2.53\lesssim a_{htt}\lesssim 2.72$. 
the value $g_{C}v$ was chosen in order to fulfill the condition $\frac{%
g_{C}^{2}}{4\pi }\lesssim 1$, which implies $g_{C}v\lesssim 0.9$ TeV. In any
case, we checked that our prediction on $R_{\gamma \gamma }$ stays almost at
the same value when the scale $g_{C}v$ is varied from $0.8$ TeV to $1$ TeV.
Considering the bounds for $\kappa $ shown in Fig.~\ref{figR}, together with
the restriction imposed by $T$ to be within its $95\%$ CL, we found that $%
M_{A}$ should have a value in a rather narrow range $1.78$~TeV$-1.9$~TeV,
while $M_{V}\lesssim 0.9M_{A}$. To arrive at this conclusion, we selected
three representative values of the axial vector mass $M_{A}$, namely at $1.78
$ TeV, $1.8$ TeV and $1.9$ TeV, and then compute the resulting $T$ and $S$
parameters. For each of these three cases, we found the corresponding values
of $M_{V}$ have to be in the ranges $1.54$ TeV $\lesssim M_{V}\lesssim $ $%
1.57$ TeV, $1.56$ TeV $\lesssim M_{V}\lesssim $ $1.59$ TeV and $1.65$ TeV $%
\lesssim M_{V}\lesssim $ $1.68$ TeV in order to have $R_{\gamma \gamma }$
within the range $0.78\lesssim R_{\gamma \gamma }\lesssim 1.55$ and the
light Higgs to have a mass $m_{h}=125.5$ GeV, without spoiling the condition 
$\frac{a_{htt}^{2}}{4\pi }\lesssim 1$.

\begin{figure*}[tbh]
\resizebox{1\textwidth}{!}{
\includegraphics{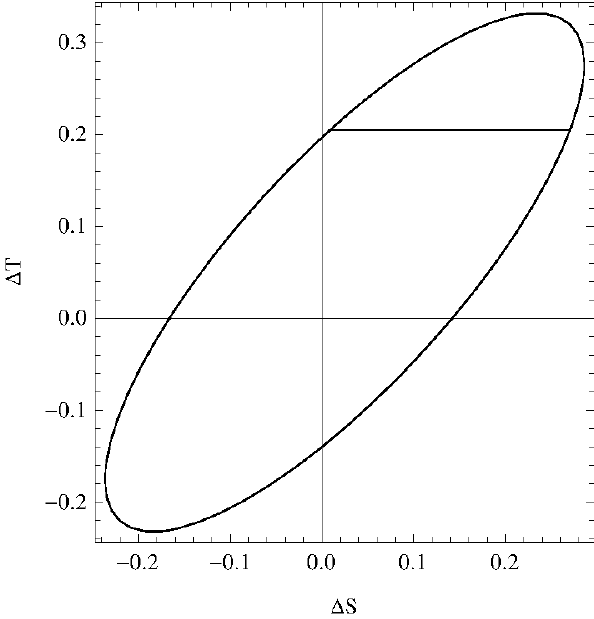}\includegraphics{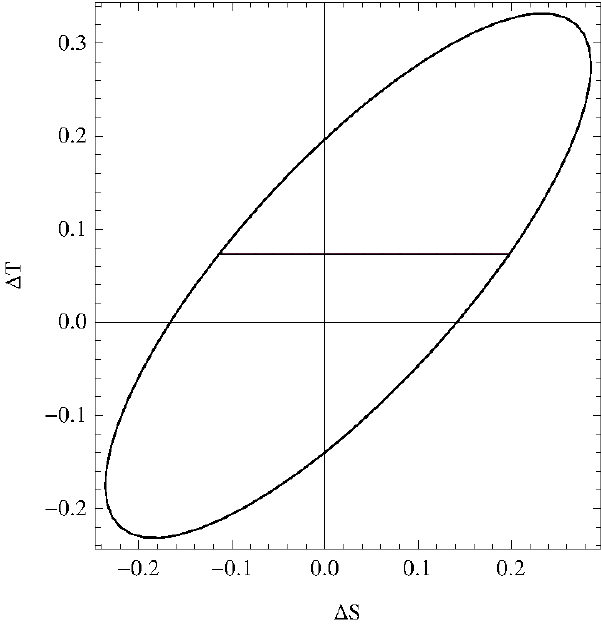}\includegraphics{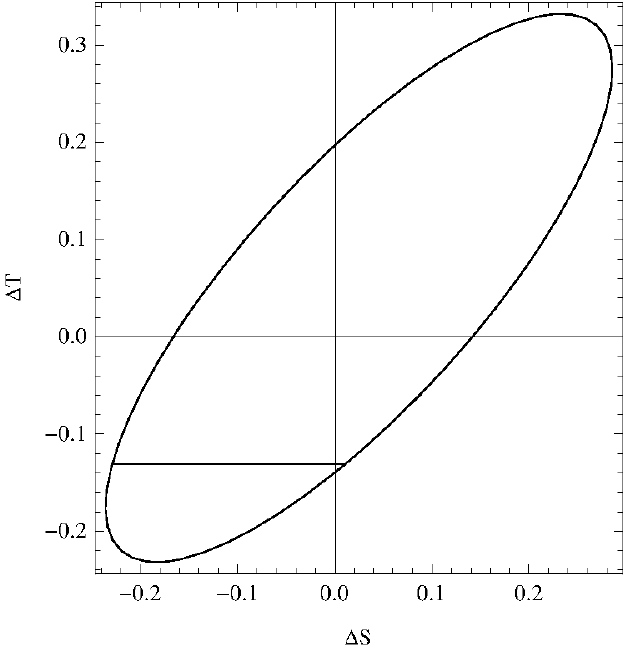}}
{\footnotesize {$M_{A}=1.78$ TeV, $M_{V}=1.55$ TeV}}\hspace{1.7cm}%
{\footnotesize {$M_{A}=1.8$ TeV, $M_{V}=1.6$ TeV}}\hspace{1.7cm}%
{\footnotesize {$M_{A}=1.9$ TeV, $M_{V}=1.7$ TeV}}\newline
{\footnotesize {\ \ \ (\ref{fig1}.a)}}\hspace{7.7cm}{\footnotesize {(\ref%
{fig1}.b)}}\hspace{5.5cm}{\footnotesize {(\ref{fig1}.c)}}\newline
\caption{The $\Delta S-\Delta T$ plane in our model with composite scalars
and vector fields  \cite{Carcamo-Hernandez:2013ypa}. The ellipses denote the experimentally allowed region at $%
95\%$CL taken from \protect\cite{GFitter}. The origin $\Delta S=\Delta T=0$
corresponds to the Standard Model value, with $m_{h}=125.5$~GeV and $%
m_{t}=176$~GeV. Figures a, b and c correspond to three different sets of
values for the masses $M_V$ and $M_A$, as indicated. The horizontal line
shows the values of $\Delta S$ and $\Delta T$ in the model, as the mixing
parameter $c_{WB}$ varies over the ranges $0.228\leq c_{WB}\leq 0.231$
(Fig.\ \protect\ref{fig1}.a), $0.208\leq c_{WB}\leq 0.212$ (Fig.\ \protect
\ref{fig1}.b), and $0.180\leq c_{WB}\leq 0.182$ (Fig.\ \protect\ref{fig1}%
.c). }
\label{fig1}
\end{figure*}

Now, continuing with the analysis of the constraints in the $\Delta T -
\Delta S$ plane, we also find that, in order to fulfill the constraint on $%
\Delta S$ as well, an additional condition must be met: for the
aforementioned range of values of $M_V$ and $M_A$, the $S$ parameter turns
out to be unacceptably large, unless a modified $W^3 - B^0$ mixing is added.
Here we introduce this mixing in terms of a coupling $c_{WB}$ [see Eq. (\ref%
{Lagrangian})]. As it is shown in Figs. 6, we find that the coupling $c_{WB}$
must be in the ranges $0.228\leq c_{WB}\leq 0.231$, $0.208\leq c_{WB}\leq
0.212$ and $0.180\leq c_{WB}\leq 0.182$ for the cases $M_A =$$1.78$ TeV, $%
1.8 $ TeV and $1.9$ TeV, respectively. In Figs.\ \ref{fig1}.a, \ref{fig1}.b
and \ref{fig1}.c we show the allowed regions for the $\Delta T$ and $\Delta S
$ parameters, for three different sets of values of $M_V$ and $M_A$. The
ellipses denote the experimentally allowed region at 95\% C.L., while the
horizontal line shows the values of $\Delta T$ and $\Delta S$ in the model,
as the mixing parameter $c_{WB}$ is varied over the specified range in each
case. As shown, $\Delta T$ does not depend on $c_{WB}$ (i.e. the line is
horizontal), while $\Delta S$ does. Moreover, the ranges for $c_{WB}$
clearly exclude the case $c_{WB}=0$, as $\Delta S$ would fall outside the
allowed region (the point would be further to the left of the corresponding
ellipse).

\section{Conclusions.}

We considered a framework of electroweak symmetry breaking without
fundamental scalars, based on an underlying dynamics that becomes strong at
a scale which we assume $\Lambda \sim 3$ TeV. %In general, below this scale there could be composite fields, bound by the strong dynamics. 
The spectrum of composite fields with masses below that scale is assumed to
consist of spin-zero and spin-one fields, and the interactions among these
particles and those of the Standard Model can be described by a $SU(2)_{L}\times SU(2)_{R}/SU(2)_{L+R}$ effective chiral Lagrangian. Specifically, the composite fields included here are two scalars, $h$ and $H$, one pseudoscalar $\eta$, a vector triplet $V_\mu^a$ and an axial vector triplet $A_\mu^a$. 
The lightest scalar, $h$, is taken to be the newly discovered state at the
LHC, with mass near 126 GeV. In this scenario, in general one must include a deviation of the Higgs-fermion coupling with respect to the SM, which is parametrized here in terms of a coupling we call $a_{hff}$. %This coupling is constrained from the requirement of having a light $125$ GeV Higgs boson mass and a two-photon signal in the range $0.78\lesssim R_{\gamma \gamma }\lesssim 1.55$ (where we use $0.78$ and $1.55$, the central values of CMS and ATLAS recent results, respectively). 
We found that our $126$ GeV Higgs boson strongly couples with the top quark
by a factor of about $2$ larger than in the Standard Model. In addition we
found that the $h\to\gamma\gamma$ rate to be consistent with the LHC
observations provided the ratio between the composite vector and axial
vector masses falls in a narrow range $M_V/M_A \sim 0.9$. We also found that
the constraints on the $T$ parameter at $95\%$C.L., together with the
previously mentioned requirement of the $h\rightarrow \gamma \gamma $ decay
rate, restrict the axial vector masses to be in the range $1.8$ TeV $%
\lesssim M_{A}\lesssim $ $1.9$ TeV. In addition, consistency with the
experimental value on the $S$ parameter requires the presence of a modified $%
W^{3}-B^{0}$ mixing, which we parametrize in terms of a coupling $c_{WB}$.
We also find that modified scalar-top quark and pseudoscalar top quark
couplings may appear, in order to have in the scalar spectrum a light $125.5$
GeV Higgs boson and heavy scalar $H$ and heavy pseudoscalar $\eta$ with
masses inside the experimental allowed range $600$ GeV $\lesssim
m_{H},m_{\eta }\lesssim 1$ TeV.

\subsection*{Acknowledgements}
This work was supported in part by Conicyt (Chile) grant ACT-119 ``Institute
for advanced studies in Science and Technology''. C.D. also received support
from Fondecyt (Chile) grant No.~1130617, and A.Z. from Fondecyt grant
No.~1120346 and Conicyt grant ACT-91 ``Southern Theoretical Physics
Laboratory''. A.E.C.H was partially supported by Fondecyt (Chile), Grant No. 11130115 and by DGIP internal Grant No. 111458. A.~E.~C.~H thanks the organizers of SILAFAE 2014 for inviting him to present this talk.
%\vspace{-0.3cm}
%\newpage
%\section{}
%\label{}



%% References

\nocite{*} 
\bibliographystyle{elsarticle-num}
\biboptions{numbers,sort&compress}
%\bibliography{martin}






% \end{thebibliography}

\end{document}